# General Model for Single and Multiple Channels WLANs with Quality of Service Support

Abdelsalam Amer, Student member, IEEE and Fayez Gebali, Senior member, IEEE

Department of Electrical and Computer Engineering, University of Victoria

Victoria BC V8W 3P6 Canada

Emails : {aamer,Fayez}@ece.uvic.ca

***Abstract***

*In this paper we develop an intergraded model for request mechanism and data transmission in the uplink phase in the presence of channel noise. This model supports quality of service. The wireless channel is prone to many impairments. Thus, certain techniques have to be developed to deliver data to the receiver. We calculated the performance parameters for single and multichannel wireless networks, like the requests throughput, data throughput and the requests acceptance probability and data acceptance probability. The proposed model is general model since it can be applied to different wireless networks such as IEEE802.11a, IEEE802.16e, CDMA operated networks and Hiperlan\2.*

## 1. Introduction

BANDWIDTH in wireless networks is in high demand. Scarce of resources and competition for access lead to degradation of the network performance. Channel utilization must be optimized by developing better medium access control (MAC) strategy and sophisticated data modulation. There are many attempts to improve the channel utilization. We have two types of networks; infrastructure and the ad hoc network. We will consider the infrastructure case in this paper where the base station (BS) coordinate the request and the data channels amongst the subscriber stations (SS). The request channels are used by the SSs to send their requests whereas the data channels are used by the SS to transmit their data. Allocating less requesting channels may lead to collision even though we get more bandwidth for data. However, allocating more requesting channels reduce the collision but affects the data bandwidth. Therefore, finding a balance point between the requesting channels and data channels is a challenge. Users require different types of wireless access or services. Centralized wireless networks have the potential of providing quality of service. The Access Point coordinate the resources amongst users. IEEE802.11a is a single channel standard and always contention happen in gaining the access. However, Point coordination function (PCF) can provide quality of service. IEEE802.11a employs carrier sense multiple access with collision avoidance (CSMA/CA) as a medium access scheme. IEEE802.16e standard can provide quality of service. In CDMA operated networks, Cai et al. in [1] studied the performance of the CDMA random access system with linear minimum mean-squared error and MF receivers and the diversity combining in fading channels. In [2] Cooper et al. investigated the problem of random-access channel performance as it pertains to wide-band code-division multiple-access (W-CDMA) wireless systems. In [3] Zhao, studied DS-CDMA with slotted aloha random access protocols. In his work he distinguished between the two stages in transmission process, the access stage and the reception stage. WLANs perform better if a cross-layer dialogue is considered and exchange of information between layers is considered. In this paper we proposed two cross-layer models to access data transmission channels. The first model is for a single class type traffic. Users compete for channel access and once they granted requests they assigned channels in the uplink to send their data. The second model, is also a cross-layer model with quality of service support. Traffic is split into two classes, high priority and low priority traffic. High priority traffic will be given more resources than the low priority traffic. The proposed models can be applied into different wireless standards. These models are generalized models since they can be applied to different wireless standards such as Hiperlan\2 [4], [5], IEEE802.11 [6] and WiMAX [7]. If we consider different number of multiple access channels then it can be applied to either Hiperlan*n*2 or WiMAX since they are multichannel standards. In that case the requesting channels will be a random access channel in Hiperlan*n*2 and it is a frequency channel in case of WiMAX. However, if only one access channel then that is a special case and applied to IEEE802.11, whereas random access channels considered as the backoff window. This model can be applied in case of CDMA operated networks where access channels are the number of codes in polls for users to compete. This paper is organized as follows; Section 2, presents the related work. Section 3 presents the network model. Single class model, its analysis and performance is presented in Section 4. Section 5 presents the quality of service support traffic model, its analysis and performance. Section 6 presets our results for both models and comments, conclusions are drawn in Section 7.

## 2. Related Work

Several cross-layer models have been proposed in WLANs [8]. Bouam in [9] proposed a cross layer design in which IEEE802.11b MAC layer used knowledge of 802.11b physical layer state to manage the channel access. Alonso et al. proposed several models in cross-layer design and QoS support using Distributed Queuing Collision Avoidance DQCA. In his work, he proposed cross-layer resource management mechanisms for voice and data traffic that combine service differentiation and opportunistic transmission [10]. In other work he also proposed a smart scheduling algorithms that operate over a near optimum MAC protocol named Distributed Queuing collision avoidance and enhance its performance [11], [12], [13], [14]. B.Walke et al. in [15], studied the performance of Hiperlan\2. In his work he presented different models for physical and data link layer. However, he did not consider the cross-layer modeling. Random access and collision reduction in Hiperlan\2 also been discussed in [16], [17], [18], [19]. In these papers, the random access channels are added based on the collision occurs in the previous MAC frame and they reduced if no access requests been issued. Also, the allocation of two slots in random access channels for each collided request reduces the MAC frame duration since the increase of random channels will effect other phases durations'. Wireless channel is prone to errors due to noise and fading. Therefore error control protocol has to be applied to deliver safe data to the receiver. Automatic-repeat-request (ARQ) techniques are used to control transmission errors. Corrupted frames have to be retransmitted in whole or only the corrupted packets in the frame. Hui Li et al. in [20] presented selective repeat and request with partial bitmap. Despite the lower overhead, still the throughput is low. Atsushi proposed PRIME-ARQ [21] that improved the throughput but lacks the flexibility to be used in different wireless standards. A.Afonso in [22] proposed an algorithm for fast retransmission and adaptive rate scheme to reduce the delay, however, the scheme reserves some bandwidth which might be not used and hence the MAC utilization is effected. Other models have been proposed but they only considered one connection or the channel error was neglected [24], [25], [26].

## 3. Network Model

In this section we show the channel utilization for the network model. Once the users sent their requests on the request channels, the successful user will be assigned certain bandwidth on the uplink. Fig. 1 shows using Time Division Duplexing (TDD) where time is broken down into frames and each frame has downlink and uplink phases. We have $k$ requesting channels and $L$ data channels. The channels could be time slots in case of the wireless networks that use the TDMA (Hiperlan 2) as their medium access. It can be frequencies for the networks that have frequency domain their medium access (WiMAX) or codes in CDMA network (3G). Users request access on the request channels $k$. The access point receives the requests and issues grants to the users. Once the users receive their allocated grants, they start sending their data. The balance point between the requesting channels $k$ and the data channels $L$ is a challenge. Increasing $k$ will reduce the collision but affects the allocated data channels and as a result degrade the throughput. On the other hand, reducing $k$ collision will increase and as a result access delay is high.

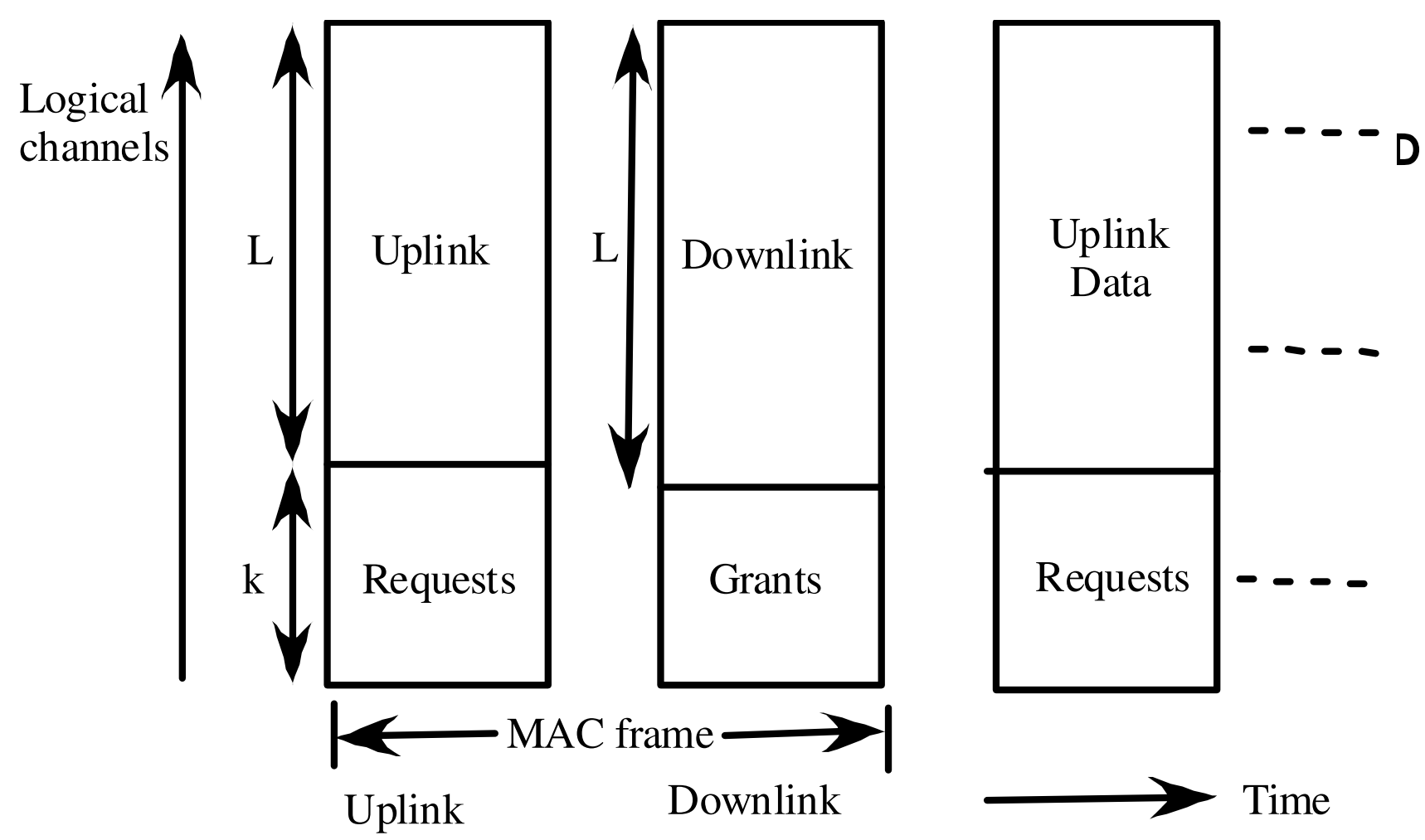

Fig. 1: Uplink and Downlink chart TDD

The uplink procedure is shown in Fig. 2. The process has six stages as numbered in Fig. 2. In the uplink phase in order for the users PDUs to be delivered they have to go through these stages; Stage 1: Users PDUs are sent by the application layers are placed in uplink queues based on their QoS criteria. Stage 2: the application layer scheduler picks up a PDU for transmission. Stage 3: MAC layer issues a request to reserve bandwidth for the scheduled PDU. Stage 4: the successfully transmitted requests from different subscriber stations are placed in the request/grant queues according to QoS criteria for both users and applications. Stage 5: the grant/application scheduler picks up which application to be sent to. Stage 6: the subscriber stations receive their grants and send their actual PDUs.

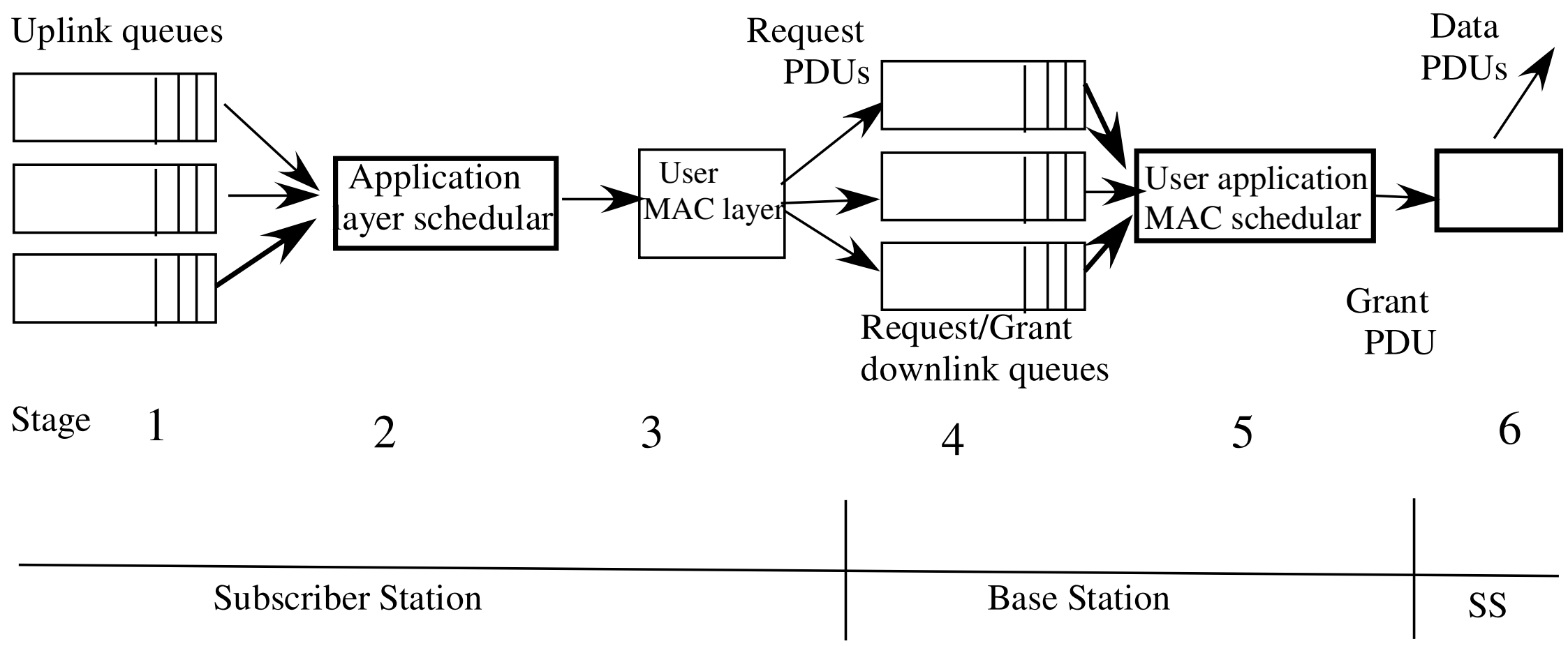

Figure 2: Uplink process

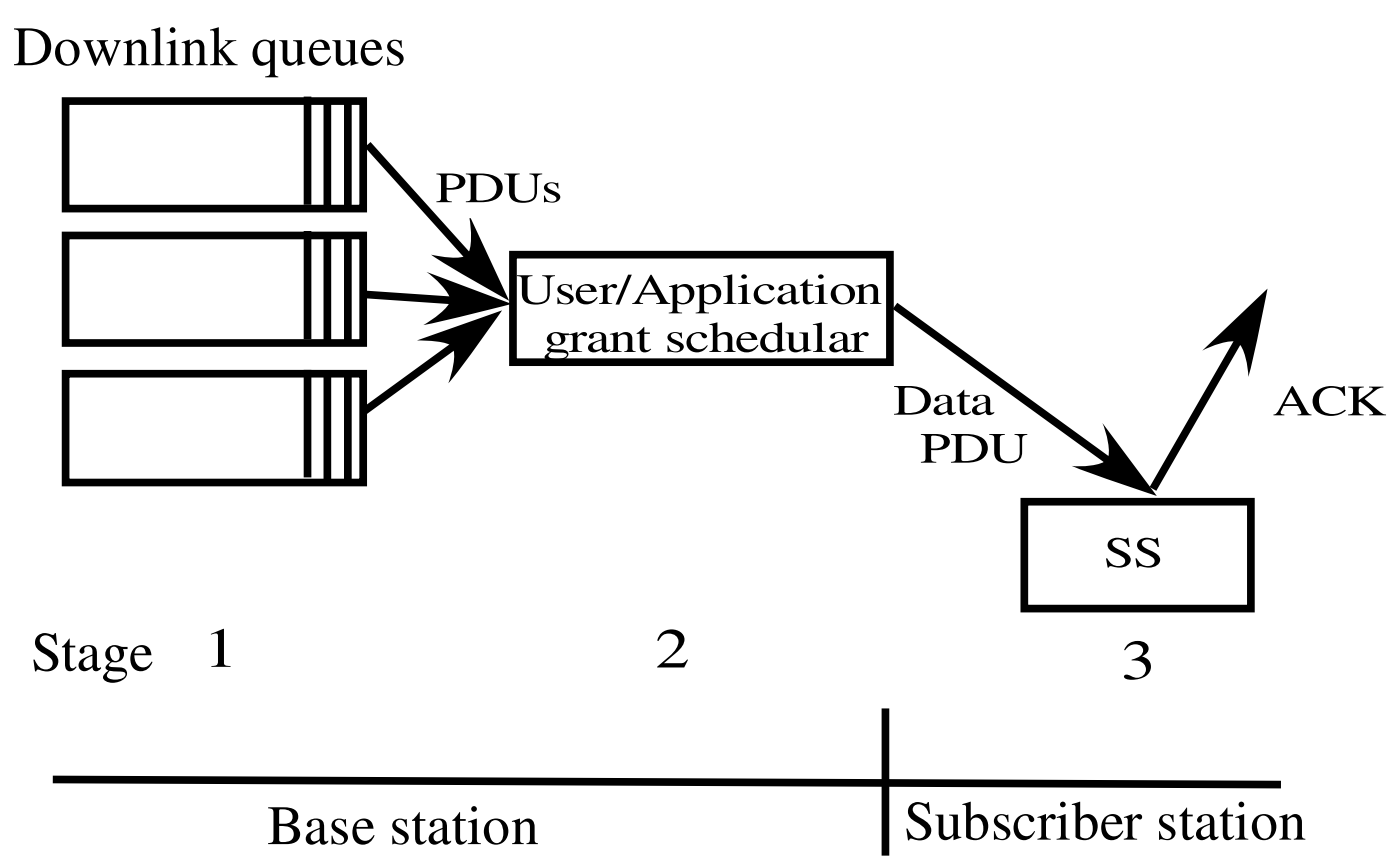

Figure 3: Downlink process

The downlink phase is shown in Fig 3. Stage 1: the successfully received requests from different subscriber stations are placed in the request/grant queues according to QoS criteria for both users and applications. Stage 2: the grant/application scheduler picks up which application to be sent to.

Stage 3: the subscriber stations receive their grants and send their actual PDUs. Once that step completed the subscriber stations send Acknowledgments.

### 1.1 Modeling Channel Error

We considered different channels (AWGN, Rayleigh fading channel and Rician channel) with different modulations scheme (BPSK and 16QAM). BPSK used as a fundamental mode in most or the wireless standard since it does not requires high SNR and usually the control data is send on this mode. The typical minimum SNR required for acceptable performance is 24*dB* [27]. We consider digitized voice with $BER = 10^{-3}$ is an acceptable error rate because it is in general can not be detected by the human ear. To maintain $BER = 10^{-3}$ in Rayleigh fading channel we need 24*dB* and it requires $SNR = 8dB$ in AWGN and 20*dB* in Rician channel. Fig. 4 shows the *SNR* versus *BER* for different channels and different modulation scheme. The figure shows the required *SNR* for these channels and modulation to obtain the targeted *BER*.

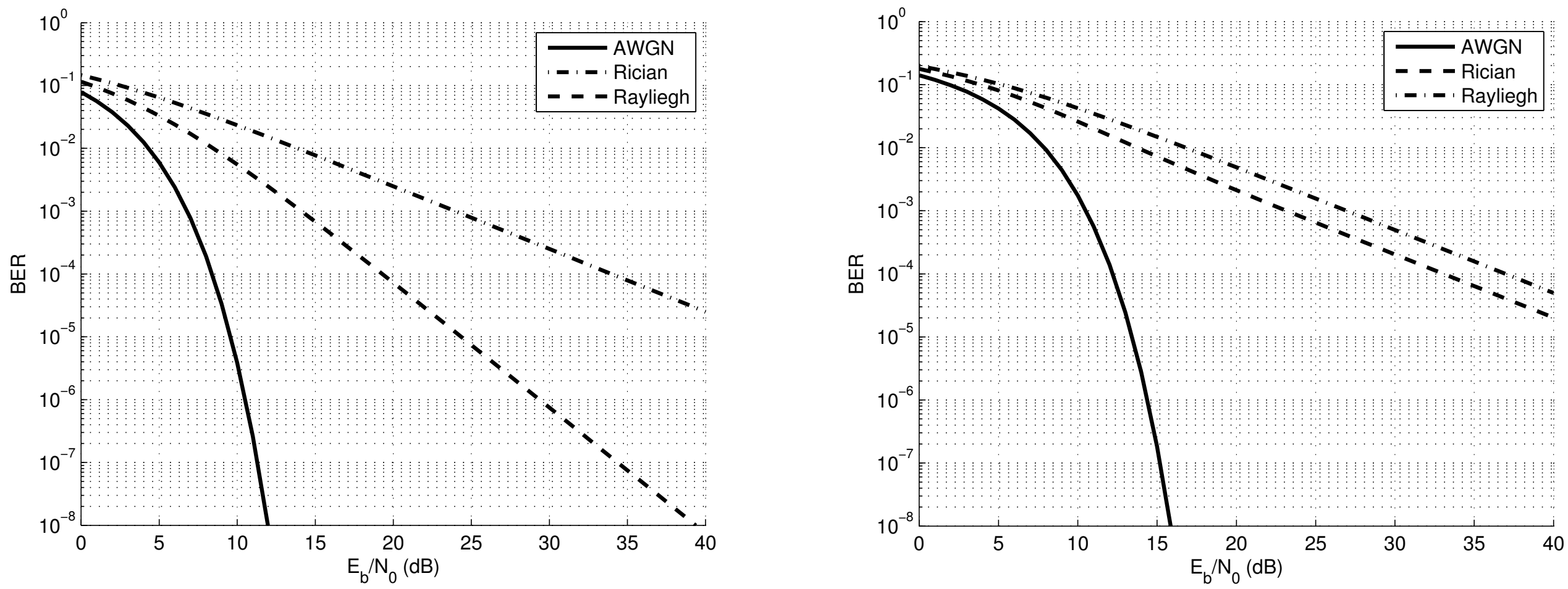

(a) SNR versus BER for BPSK (b) SNR versus BER for 16QAM

Fig. 4: SNR versus BER for different modulation and channels

## 4. Single Class Model

In this section we will present our proposed single class model where all users given similar priority

### 4.1 Model Analysis

In this section, we will show the single class model analysis. We assume that *N* users try to request access on the random requesting channels. The number of request channels is assumed to be *k*.
In order to analyze the system behavior, some assumptions are made ;
1) The probability that a user issues a request is *a*.
2) The probability a user chooses a particular reservation channel is 1/*k*.
3) A collided user retransmits with probability *c*.
4) The traffic is calculated in one radio cell. No outside traffic is considered.
5) The average length of a packet is *nb* bits.

6) The probability that the transmitted packet contained error is *e*.
7) The feedback channel is error free.
8) The sender will keep sending a packet *n* times.
The error control protocol states as shown in Fig. 5 represented by *st*0 until *stn* have the Following properties:
1) State *sti* indicates that the SS is retransmitting the frame for the $i^{th}$ time whereas, state *st*0 Indicates error-free transmission.
2) The forward channel has random noise and the probability that a bit will be received in error is $\varepsilon$, (*BER*).
3) The number of transmission states is $n + 1$.
4) The time step is taken equal to the sum of transmission delay (time required to send a frame) and round trip delay (time required for frame propagation and reception of acknowledgment).

A Subscriber Station (SS) that has data to send issues a request on the random requesting channels. Contention may occur if two or more SSs choose the same requesting channel. A user could be in one of three states; *transmit* state, if a single request received or *collide* state, if two or more SSs issues a request on the same channel or *idle* state if there is no request has been received. Fig. 5 shows the Markov chain state diagram for a user with error control. The collided users adapts a constant probability backoff in which the collided users retransmit with a probability.

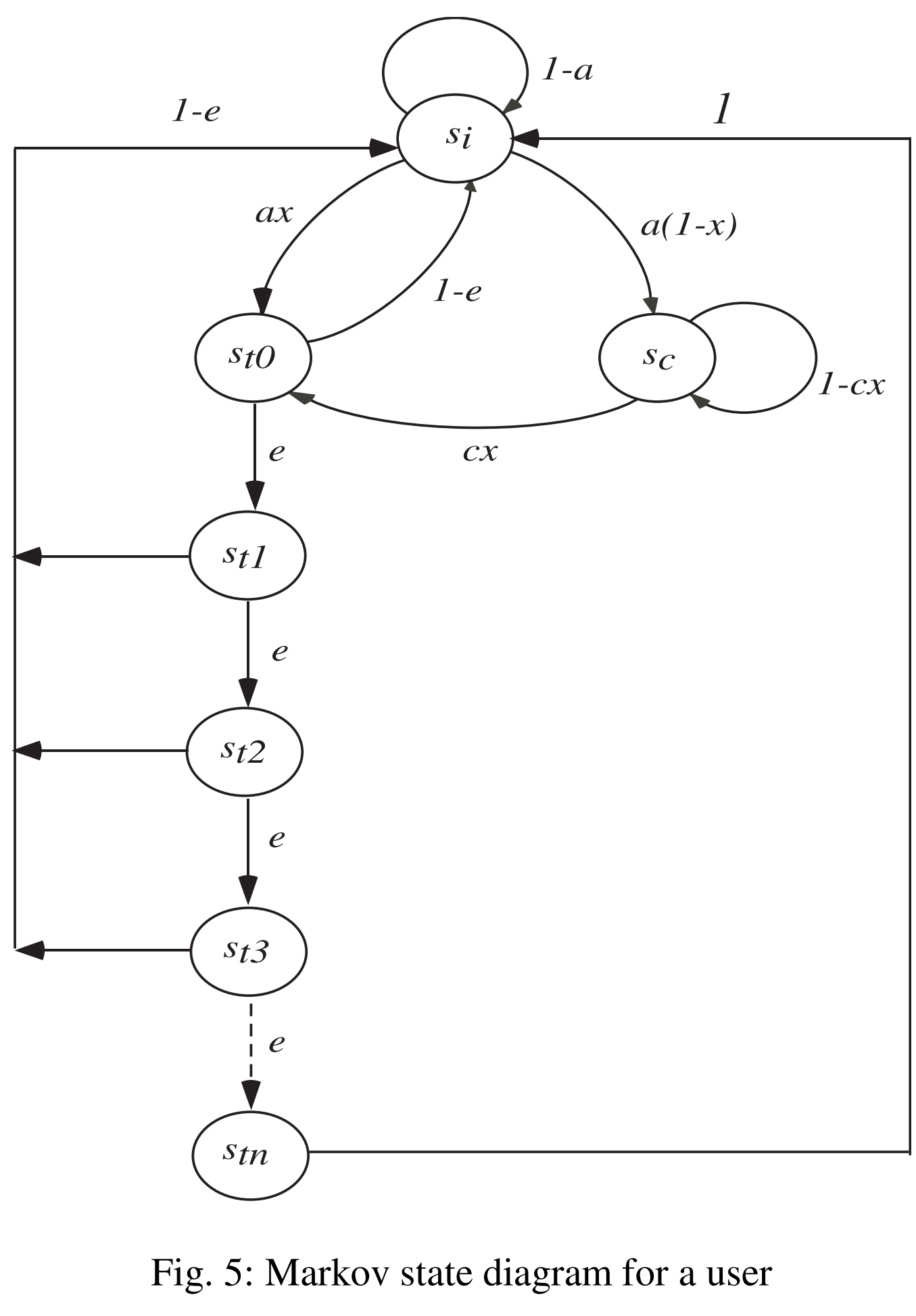

Fig. 5: Markov state diagram for a user

The error is calculated by;

$$e = 1 - (1 - \varepsilon)^{nb} \qquad (1)$$

where *nb* is the number of bits in a message. *x* is the probability that a user successfully accesses one of the free channels and it is given by;

$$x = (1 - \tfrac{1}{k})^{N_{ave} - 1} \qquad (2)$$

where $N_{ave}$ is the average number of active users;

$$N_{ave} = N(as_i + cs_c) \qquad (3)$$

$$y = 1 - x \qquad (4)$$

*y* is the probability that a user selects a busy channel.
A discrete-time Markov chain is characterized by the transition matrix P which can be obtained from the state diagram and the state vector s [28]. The state vector s for the user is organized as follows;

$$s = [s_i \quad s_c \quad s_{t0} \quad s_{t1} \quad s_{t2} \quad \cdots \quad s_{tn}]^t \qquad (5)$$

where *si* is the probability that the user is in the *idle* state, *st* is the probability that the user is in the *transmit* state and *sc* is the probability that the user is in the *collide* state. The SS will keep sending the packet if there is no acknowledgment is received (i.e the packet sent with an error probability *e*) *n* times. When a packet is correctly received the SS goes to *idle* state with probability 1-*e*.
At equilibrium, the distribution vector elements are obtained by solving the following two equations [28];

$$Ps = s \qquad (6)$$

$$\sum s_j = 1 \qquad (7)$$

Where $j \in \{i, t1, t2, t3, \cdots tn, c\}$

From Eqs.(6) and (7) we can find the state vector elements at equilibrium

$$s_i = \frac{1}{D_{n2}}$$

$$s_c = \frac{a(1-x)}{D_{n2}}$$

$$s_{tj} = \frac{B}{D_{n2}} \sum_{j=0}^{n-1} e^j \qquad (8)$$

Where B

$$B = ax[1 + c(1-x)]$$

And $D_{n2}$ is

$$D_{n2} = 1 + B + eB + e^2B + e^3B + \cdots + e^{n-1}B + a(1-x) \qquad (9)$$

### 4.2 Performance for the single class model

We study the performance of this model in this subsection. We applied one backoff strategy model (Constant backoff probability model) as an example. The average number of retransmission: The average number of retransmissions due to error for a packet using Stop and Wait(SW) protocol is given by [30] , [31]:

$$N_t = s_1 + 2s_2 + 3s_3 + \cdots + ns_n$$
$$= eB + 2e^2B + 3e^3B + \cdots + ne^nB$$
$$= \sum_{i=1}^{n} ie^iB \quad transmissions / packet \tag{10}$$

The efficiency: The efficiency is defined as the total number of transmission which indicates the first retransmission plus the average number of retransmission and it is given by:

$$\eta = \tfrac{1}{1+N_t} \tag{11}$$

Throughput: The throughput is obtained from the following equation:

$$Th = \min(Ns_t, k) \tag{12}$$

Acceptance probability: The acceptance probability is defined as the ratio between the throughput and the offered load [28]:

$$p_a = \frac{Th}{Na} \tag{13}$$

Access delay: The access delay (*D*) is the average number of access attempts made by the SSs before they are successfully granted a channel. It is defined as;

$$D = \sum_{i=0}^{\infty} i(1-p_a)^i p_a$$
$$= \frac{1-p_a}{p_a} \tag{14}$$

Energy: The average energy *Ea* required to transmit a request successfully can be calculated as follows [29];

$$E_a = E_0 \sum_{i=0}^{\infty} (i+1)(1-p_a)^i P_a$$
$$= \frac{E_a}{p_a}$$
$$E_a[dB] = -10\log(p_a) \tag{15}$$

Where *E*0 is the energy required to transmit a request once.

Uplink channel utilization: Equation (16) calculates the data uplink channel utilization

$$\eta_u = \frac{\min\{L, Ns_t\}}{L} \tag{16}$$

Net acceptance probability: This equation is to find the net acceptance probability for the data channel

$$Pa(net) = \begin{cases} p_a & NTh < L; \\ p_a \dfrac{L}{Ns_t} & NTh > L \end{cases} \tag{17}$$

Where $L$ is the number of data channels, $N$ is the number of users, and $s_t$ is success probability extracted from the state vector.
In the next section we will extend our model to support quality of service.

## 5. Quality of Service Support Model

In the two-class priority model (Quality of Service support model), the total number of random channels is split into two groups $k1$ and $k2$, where $k2 < k1$ as shown in Fig. 6. Traffic is classified into two classes, high priority class and low priority class. From Fig. 6, high priority class traffic users compete for access on $k1$ channels and low priority class traffic users compete for access on $k2$ channels. Also data channels are divided into $L1$ for high priority traffic and $L2$ for low priority traffic. Fig. 6 shows the Quality of service model chart TDD. In the uplink phase where we have $L$ data channels. The data channels are split into classes where each class is assigned number of channels. The assignment of channels is based on the priority of the traffic class.

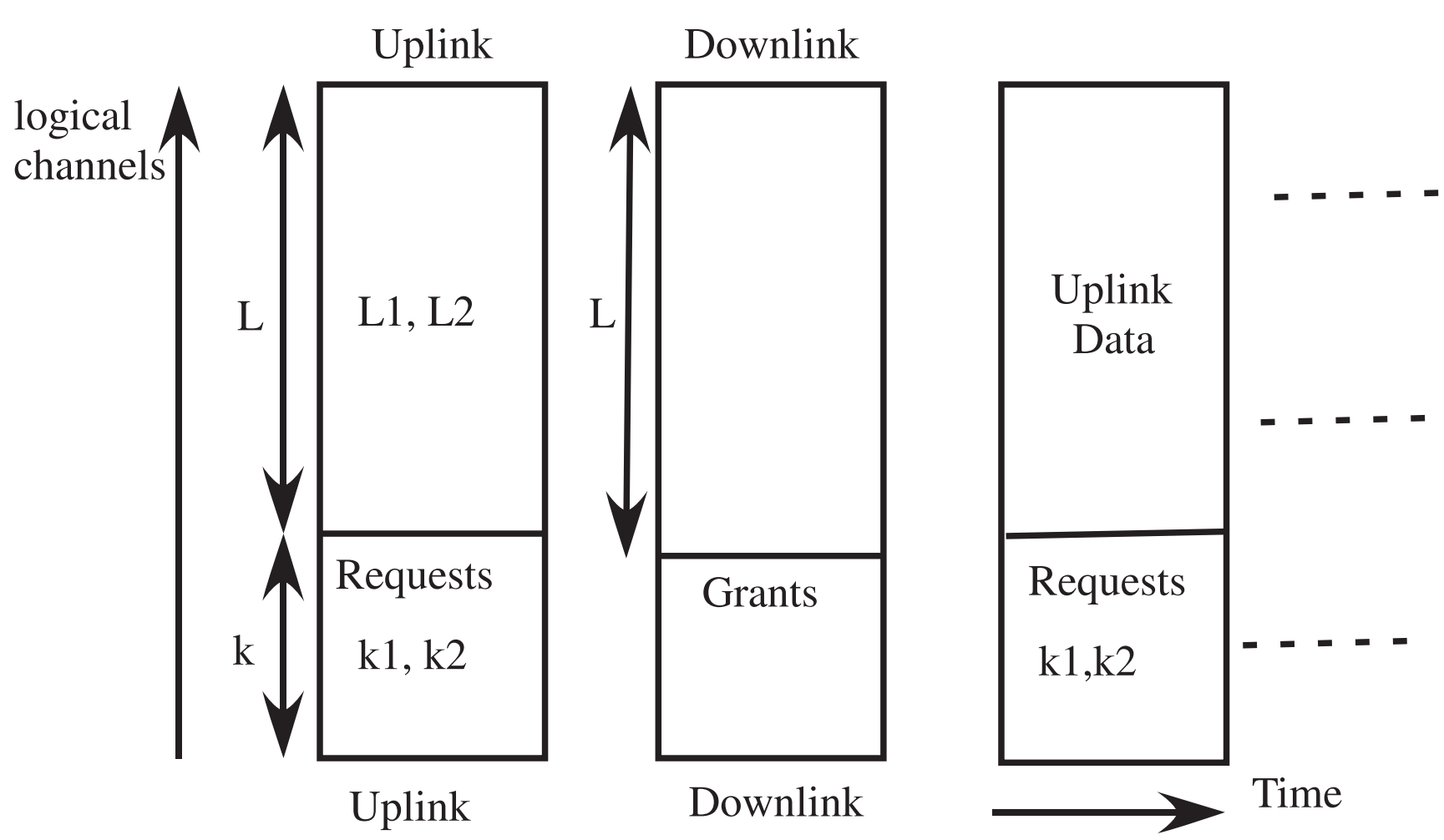

Fig. 6: Uplink and downlink quality of service chart TDD

### 5.1 Quality of Service Model Analysis

In this subsection we will show the quality of service model analysis. We assume that we have $N$ users try to request access to send their data in the uplink phase. An arriving packet belongs to high priority traffic class with probability $l$ and belongs to low priority class with probability $1 - l$. Contention may occur if two or more SSs choose the same channel. A SS could be in one of the three states; *transmit* state, if a single request is received or *collide* state, if two or more SSs issue requests on the same channel or *idle* state if there is no request has been received. Fig. 7 shows the Markov chain state diagram for the user. There are two classes of users. In order to analyze the system behaviour some assumptions are made;

1) The probability that a user issues a request is $a$
2) The probability that a user from high priority class chooses a particular channel is $1/k1$ and

$1/k2$ for low priority class.

3) The collided user from high priority class retransmits with probability $c1$ and from low priority Class is $c2$.
4) The probability that the transmitted packet contained error is $e$.
5) The forward channel has random noise and the probability that a bit will be received in error is $\varepsilon$ ($BER$).
6) The feedback channel is error free.

The error control states as shown in Fig. 7 represented by $st1$ until $stn$ have the following properties:

1) State $sti$ indicates that the SS is retransmitting the packet for the $i^{th}$ time.
2) The number of transmissions states is $n$.
3) The time step is taken equal to the sum of transmission delay (time required to send a packet) and round trip delay (time required for packet propagation and reception of acknowledgment) .

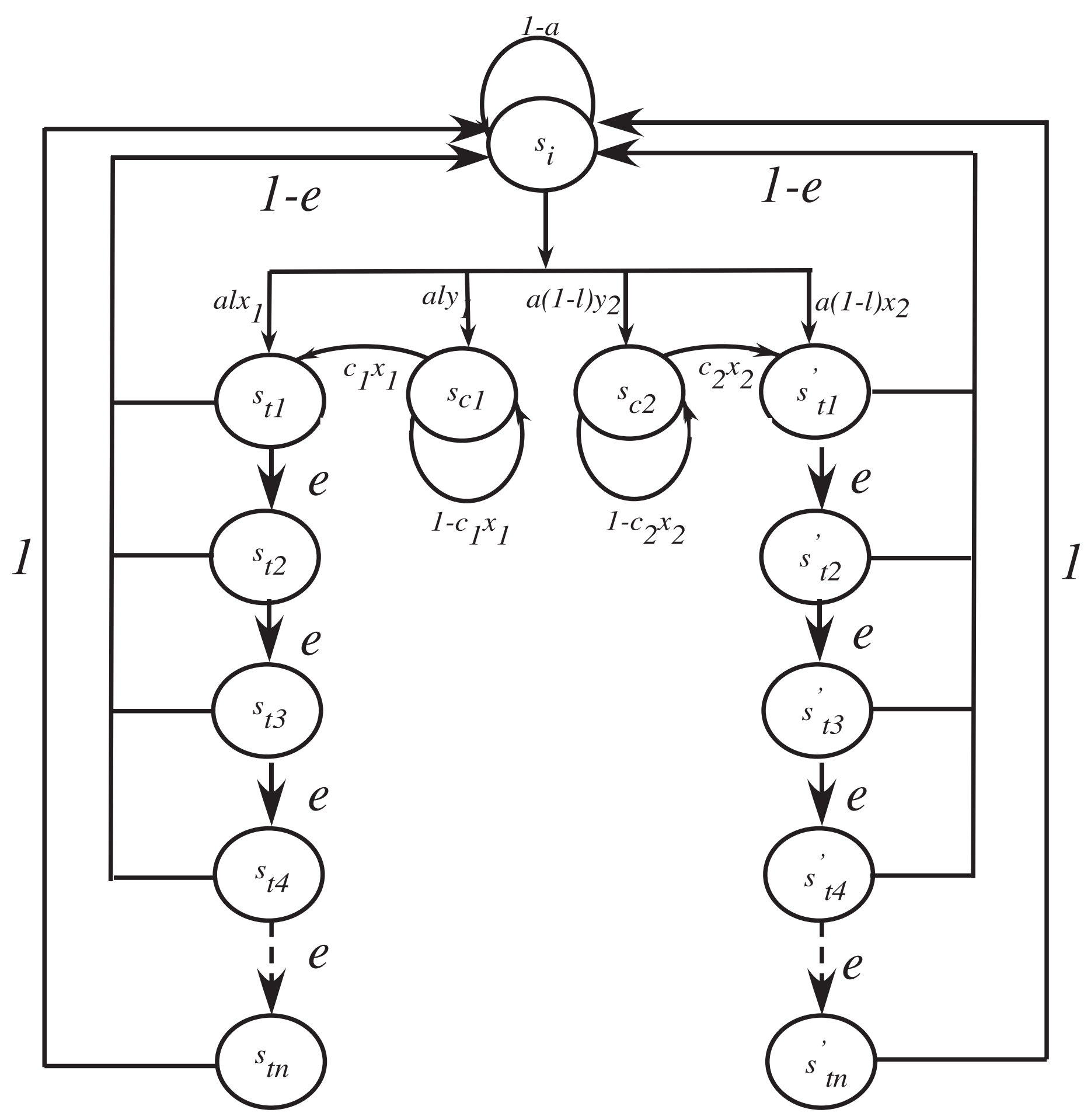

Fig. 7: Markov state diagram for users of two-class

The error is calculated by;

$$e = 1 - (1-\varepsilon)^{nb} \quad (18)$$

Where nb is the number of bits.
From Fig. 7, $x1$ is the probability that a user from class one successfully accesses one of the $k1$ random channels. $x_1$ is given by:

$$x_1 = (1 - \tfrac{1}{k1})^{N_{1a}-1} \quad (19)$$

where $N_{1a}$ is the average number of active users from high priority class (class one) and it is calculated by:

$$N_{1a} = N(las_i + c_1 s_{c1}) \quad (20)$$

The probability that the user from class one (high priority) selected one of the collided channels is given by:

$$y_1 = 1 - x_1 \quad (21)$$

$x_2$ is the probability that a user from low priority class (two) successfully accesses one of the $k2$ random channels. $x_2$ is given by:

$$x_2 = (1 - \tfrac{1}{k2})^{N_{2a}-1} \quad (22)$$

where $N2a$ is the average number of active users from low priority class (class two) and it is calculated by:

$$N_{2a} = N[(1-l)as_i + c_2 s_2)] \quad (23)$$

The probability that the user from class two (low priority) selected one of the collided channels is given by:

$$y_2 = 1 - x_2 \quad (24)$$

The allocated channels for the two classes are allocated by:

$$k1 = \left\lfloor \frac{lm}{lm + (1-l)} k_{\max} \right\rfloor \quad (25)$$

and the allocated channels for the second class are calculated by:

$$k2 = k_{\max} - k1 \quad (26)$$

where $m$ is the weight factor to determine the number of channels allocated to both classes and $k_{max}$ is the maximum number of allocated random channels.
A discrete-time Markov chain is characterized by the transition matrix P and the state vector s [28]. The state vector s for the user is organized as:

$$s = [s_i \quad s_{c1} \quad s_{t1} \quad s_{t2} \quad s_{t3} \quad \mathrm{L} \quad s_{tn} \quad s_{c2} \quad s_{t1}' \quad s_{t2}' \quad s_{t3}' \quad \mathrm{L} \quad s_{tn}']^t \quad (27)$$

The SS will keep sending the packet contained if there is no acknowledgment is received (i.e the packet sent with an error probability $e$). When a packet is correctly received the SS goes to *idle* state with probability 1-$e$. The corresponding state transition matrix for the user which is extracted from the state transition diagram shown in Fig. 7
At equilibrium, the distribution vector elements are obtained by solving the following two equations [28]:

$$Ps = s \tag{28}$$

$$\sum s_j = 1 \tag{29}$$

Where $j \in \{i, c1, t1, t2, t3, \cdots tn, c2, t1', t2', t3', \cdots tn'\}$

by solving Eqs. (28) and (29) we can find the state vector elements at equilibrium

$$s_i = \frac{1}{D_{n1}}$$

$$s_{c1} = \frac{1}{D_{n1}} aly_1$$

$$s_{tj} = \frac{B_1}{D_{n1}} \sum_{j=0}^{n-1} e^j$$

$$s_{c2} = \frac{1}{D_{n1}} a(1-l)y_2$$

$$s'_{tj} = \frac{B_2}{D_{n1}} \sum_{j=0}^{n-1} e^j$$

where $D_{n1}$ is

$$D_{n1} = [1 + B_1 + eB_1 + e^2 B_1 + e^3 B_1 + \cdots + e^{n-1} B_1 + aly1 + B_2 + eB_2 + e^2 B_2 + e^3 B_2 + \cdots + e^{n-1} B_2 + a(1-l)y_2] \tag{30}$$

and

$$B_1 = alx_1 + alc_1 x_1 y_1$$

$$B_2 = a(1-l)x_2 + c_2 x_2 a(1-l)y_2 \tag{31}$$

### 5.2 Quality of Service Model performance

Based on the system analysis, we study the performance of the quality of service support model
Throughput: The throughput for both classes can be calculated as follows:

$$Th_i = \min(Ns_{ti}, k_i) \qquad i \in \{1,2\} \tag{32}$$

Acceptance probability: The packet acceptance probability is defined as the ratio between the throughput and the offered load [28]:

$$p_{a1} = \frac{Th_1}{lNa} \tag{33}$$

$$p_{a2} = \frac{Th_2}{(1-l)Na} \tag{34}$$

Access delay: The access delay ($D_{1,2}$) is the average number of access attempts made by the SS before it is successfully granted a random access channel. It is defined as:

$$D_i = \frac{1 - p_{ai}}{p_{ai}} \qquad i \in \{1,2\} \tag{35}$$

Average energy: The average energy $E_{a1,a2}$ required to transmit a request successfully can be calculated as follows [29]:

$$Ea1 = \frac{E_0}{p_{a1}}$$

where $E0$ is the energy required to transmit a request once. Normalizing relative to $E0$, average energy in $dB$ is given by:

$$E_{a1}[dB] = -10\log(p_{a1}) \quad (36)$$

$$E_{a2}[dB] = -10\log(p_{a2}) \quad (37)$$

Average number of retransmissions: The performance of the error control protocol is measured by the average number of retransmission of a packet and the efficiency. The average number of retransmission for each class is given by [28], [32] , [33]:

$$\begin{aligned} N_{t1} &= s_{t1} + 2s_{t2} + 3s_{t3} + \cdots + ns_{tn} \\ &= eB_1 + 2e^2B_1 + 3e^3B_1 + \cdots + ne^nB_1 \\ &= \sum_{i=1}^{n} ie^iB_1 \quad transmissions/packet \end{aligned} \quad (38)$$

$$\begin{aligned} N_{t2} &= s'_{t1} + 2s'_{t2} + 3s'_{t3} + \cdots + ns'_{tn} \\ &= eB_2 + 2e^2B_2 + 3e^3B_2 + \cdots + ne^nB_2 \\ &= \sum_{i=1}^{n} ie^iB_2 \quad transmissions/packet \end{aligned} \quad (39)$$

Efficiency: The efficiency for each class is calculated by [28]:

$$\eta_i = \frac{1}{1+N_{ti}} \qquad i \in \{1,2\} \quad (40)$$

For the error free channel $e = 0$ and the average number of retransmissions is 0. That means the packet is sent only once for a successful transmission.

Channel utilization: This equation is to calculate the channel utilization for both classes

$$\eta_{ui} = \frac{\min\{L_i, Ns_{ti}\}}{L_i} \qquad i \in \{1,2\} \quad (41)$$

where $L_i$ are the allocated data channels for both classes.

Net acceptance probability: The net acceptance probabilities for data channels for both classes can be calculated by;

$$p_{a(net)1} \begin{cases} p_{a1} & NTh1 < L_1; \\ p_{a1}\dfrac{L_1}{Ns_{t1}} & NTh1 > L_1 \end{cases} \quad (42)$$

And for class two traffic

$$p_{a(net)2} \begin{cases} p_{a2} & NTh2 < L_2; \\ p_{a2}\dfrac{L_2}{Ns_{t2}} & NTh2 > L_2 \end{cases} \quad (43)$$

# 6 Results

In the performance, we used, $N = 50$, number of packets = 500, $BER = 10^{-3}$, constant backoff probability is assumed with retransmission probability $c = 0.75$ and $k = 25$. Fig. 8 and 9 show the obtained results for the single class model. Fig. 8, shows the requests throughput of the single class. The throughput is increasing with the incoming traffic and starts to go down in the heavy traffic which is natural since the resources are limited and the collided users are retransmitting. Fig. 9a shows the channel utilization when we vary $L = \{1, 5, 10, 15\}$.

When $L = 1$ which is a special case for IEEE802.11a we notice that the channel is fully utilized. When we increase $L$ the channel utilization is going down which is the case for multiple channel access WLANs and better chance for data transmission is the channels not fully utilized.

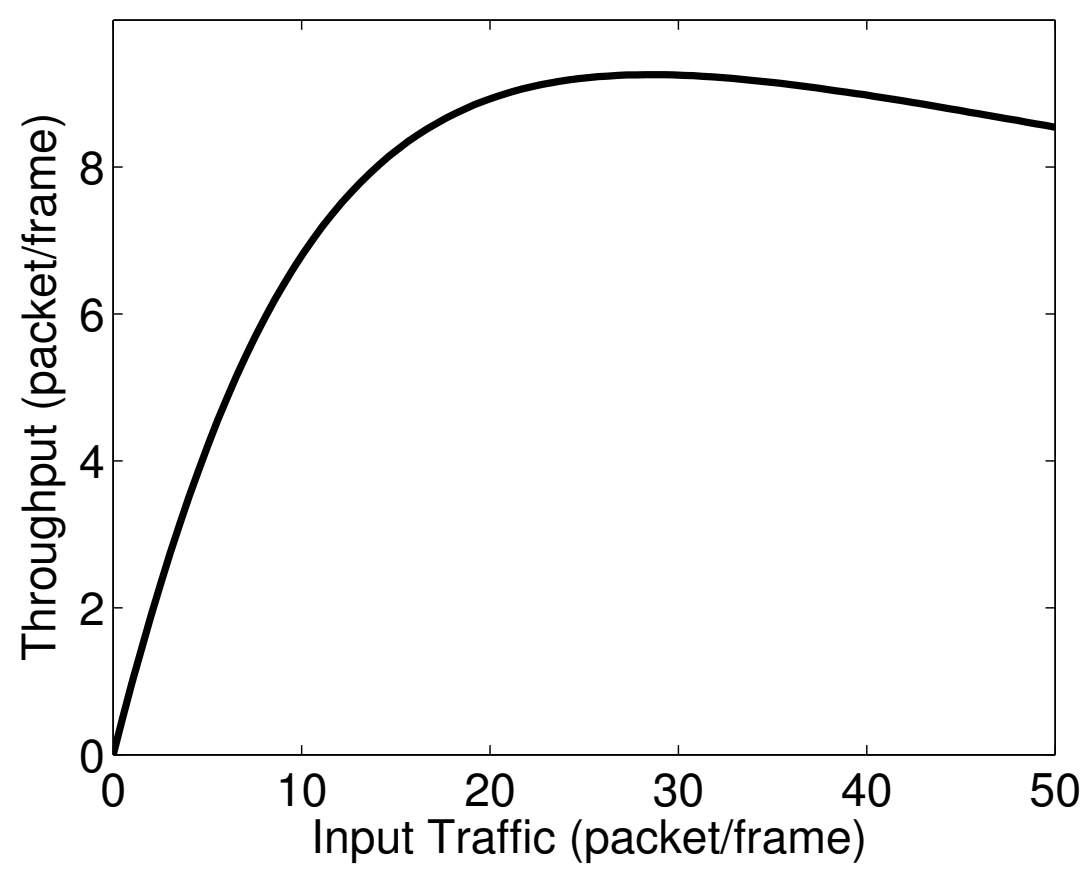

Fig. 8: Requests throughput for single class model

Fig. 9a, shows the net acceptance probability for the single class model. When $L = 1$, 802.11a case, the net acceptance probability is lower compared to the other wireless multichannel standards. Also, as the number of data channels increases the net acceptance probability increases until it gets to the request acceptance probability. That means, all granted requests get their data transmitted.

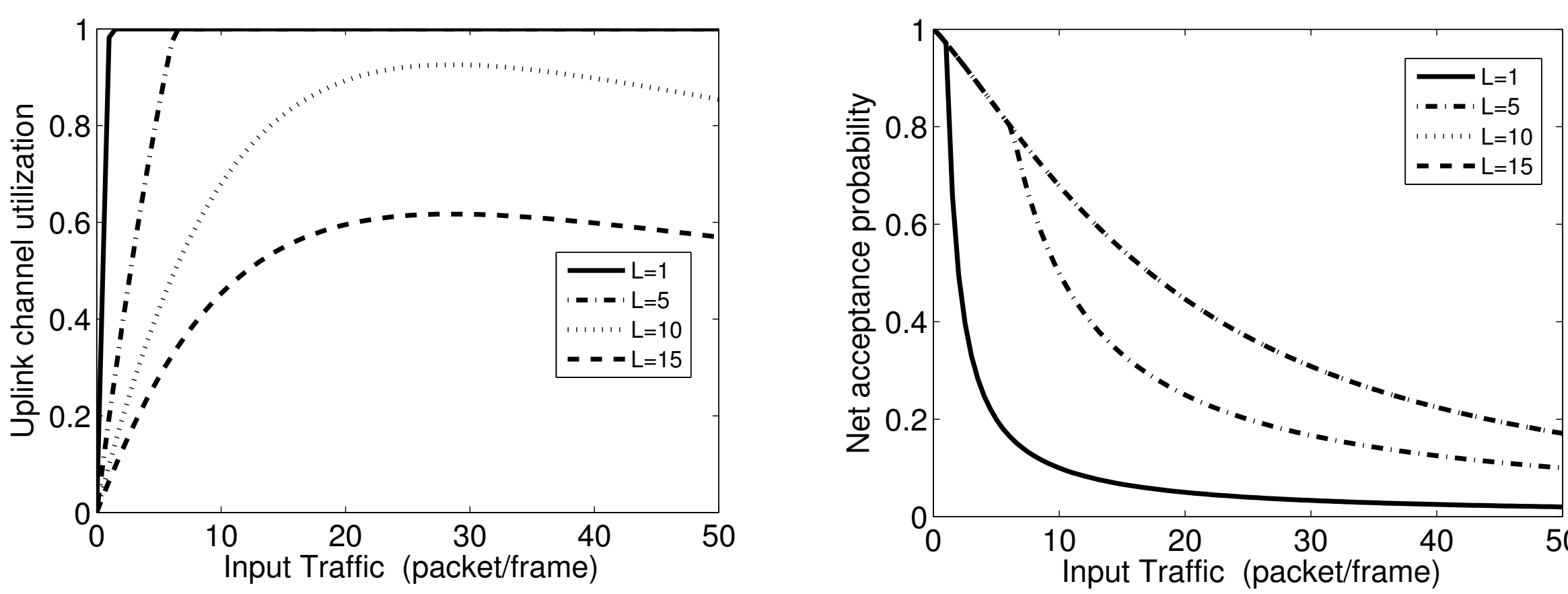

(a) Uplink channel utilization versus input traffic (b) Net acceptance probability versus input traffic

Fig. 9: Performance for single class model

Fig. 10 shows the throughput request for the quality of service support model. we assume that we have, $N = 50$, $kmax = 20$, $L1 = L2 = \{1, 5, 10\}$, $c1 = c2 = 0.75$, $m = 2$, $l = 0{:}75$, $BER = 10^{-3}$ and $NP = 500$. Fig. 10a, shows the throughput request for the quality of service model. We notice that quality of service guarantee for all types of traffic for the high priority class traffic.

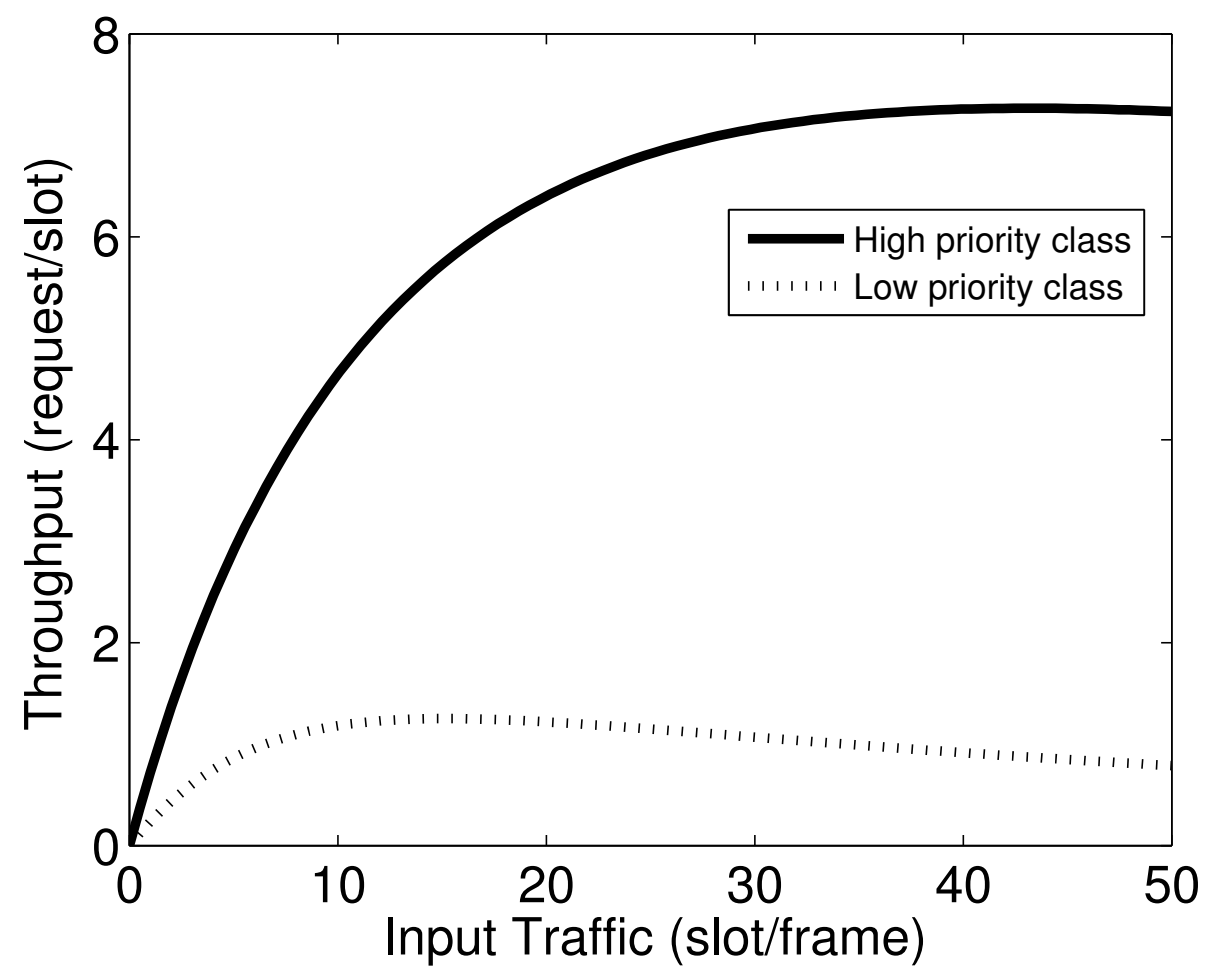

Fig. 10: Requests throughput for QoS model

Fig. 11, shows the channel utilization. From the figure, when $L1,L2 = 1$, which is a special case for IEEE802.11a, where we have one channel, we notice that the channel is fully utilized with class one users when the input traffic is low and kept fully utilized. When $L1 = L2 = 5, 10$ the channel utilization is decreased and that is a better chance for data to be transmitted. That case is for the mutliple channel WLANs. We also guarantee quality of service in the high priority class traffic as can be seen from Fig. 11a and Fig. 11b. Fig. 12, shows the net acceptance probability for both classes. When the $L1,L2$ are small the net acceptance probability is low, however, the acceptance probability is getting higher as the input traffic increases. Furthermore, the net acceptance probability of the low priority class is better when $L1,L2 = 1$ the reason is that more traffic is coming from higher priority class than low priority class. However, when $L1,L2$ increases the net acceptance probability of higher priority class is better since the allocated resources are fully used. We also show that as the number of access data channels increases the net acceptance probability is similar to the access request acceptance probability. In that case, all granted requests get their data transmitted.

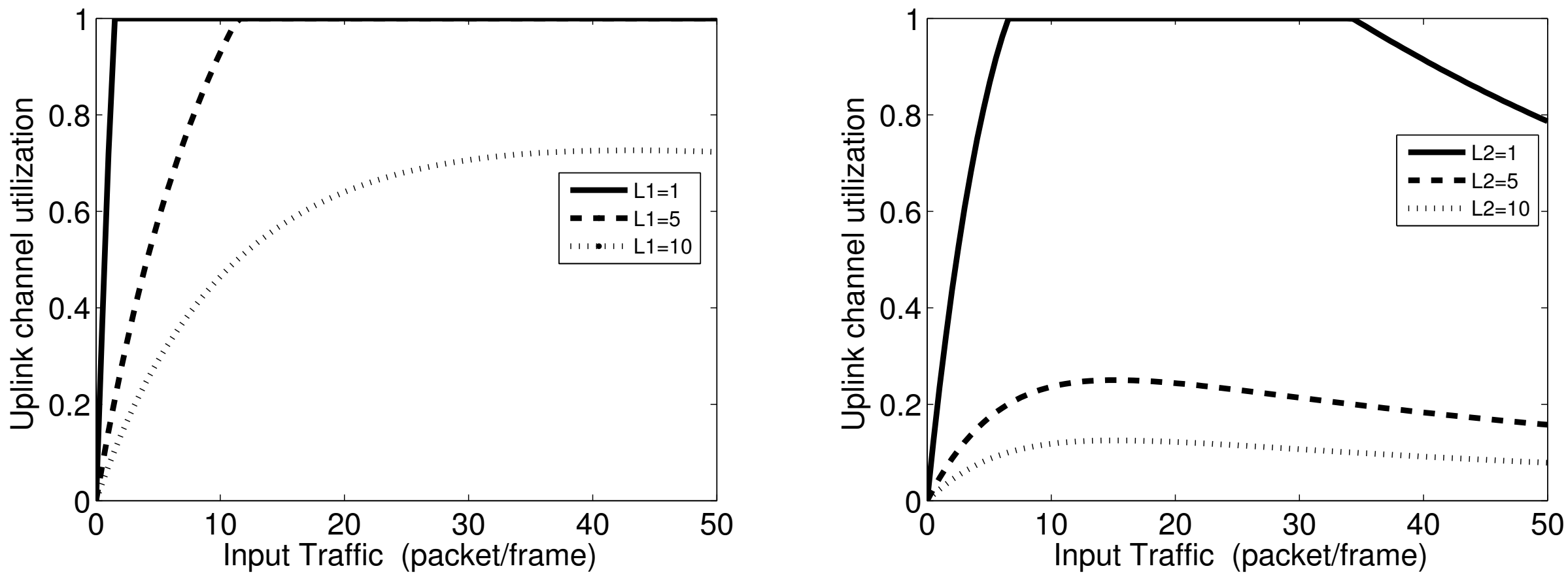

(a) Class one uplink utilization (b) Class two uplink utilization

Fig. 11: Uplink channel utilization for QoS support model

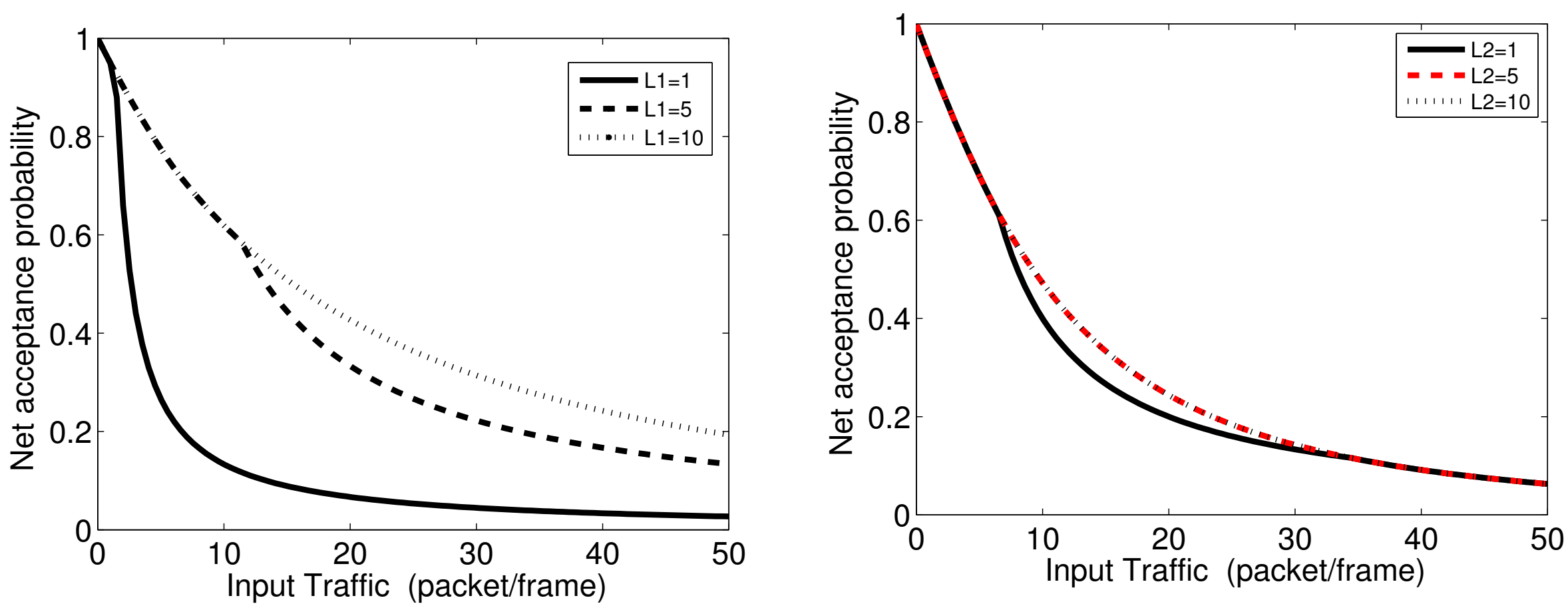

(a) Class one net acceptance probability (b) Class two net acceptance probability

Fig. 12: Net acceptance probability for QoS support model

# 7 Conclusions

In this paper we developed two models for the channel utilization, one class and quality of service support models. We also studied the net acceptance probability for these two models. In the single class model the channel is fully utilized when we have only one data channel which is a special case for IEEE802.11a. However, the utilization is lower when we have more data channels. Furthermore, the net acceptance of one class model is lower when there is only one data channel. However, as the number of data channels increase the net acceptance probability improved. In the quality of service model, the uplink channel utilization reaches full channel utilization when we have only one channel which is a special case for IEEE802.11a and the utilization starts to go down as we have more channels. We also assured that high priority class get better performance, so quality of service is assured for it. In a similar way we assured the net acceptance probability for the high priority class in all types of traffic. These two models can be applied into different wireless standards. In the quality of service support model high priority get better access chance and the low priority is not ignored when high priority class requesting access.